\documentclass[12pt]{article}

\usepackage{amssymb,epsfig,psfig}

\begin{document}

\begin{center}
\baselineskip=24pt

{\Large \bf Study and suppression of anomalous fast events in inorganic 
scintillators for dark matter searches}
\vspace{0.5cm}

\baselineskip=18pt

{\large V. A. Kudryavtsev\footnote{Corresponding author, 
e-mail: v.kudryavtsev@sheffield.ac.uk}, N. J. C. Spooner, P. K. Lightfoot, \\
J. W. Roberts\footnote{Now at the Rutherford Appleton Laboratory}, 
M. J. Lehner\footnote{Now at the University of Pennsylvania, Philadelphia}, 
T. Gamble, M. J. Carson, \\
T. B. Lawson, R. L\"uscher\footnote{Now at the Imperial College}, 
J. E. McMillan, B. Morgan, \\
S. M. Paling, M. Robinson, D. R. Tovey}

\vspace{0.2cm}
{\it Department of Physics and Astronomy, 
University of Sheffield, Sheffield S3 7RH, UK}

\vspace{0.2cm}
{\large N. J. T. Smith, P. F. Smith, G. J. Alner, S. P. Hart, \\
J. D. Lewin, R. M. Preece}

\vspace{0.2cm}
{\it Particle Physics Department, 
Rutherford Appleton Laboratory, Chilton, Oxon \\
OX11 0QX, UK}

\vspace{0.2cm}
{\large T. J. Sumner, W. G. Jones, J. J. Quenby, B. Ahmed, A. Bewick, \\
D. Davidge, J. V. Dawson, A. S. Howard, 
I. Ivaniouchenkov\footnote{Now at the Rutherford Appleton Laboratory},\\
M. K. Joshi, V. Lebedenko, I. Liubarsky}

\vspace{0.2cm}
{\it Blackett Laboratory, Imperial College of Science, 
Technology and Medicine, London SW7 2BZ, UK}

\vspace{0.2cm}
{\large J. C. Barton}

\vspace{0.2cm}
{\it Department of Physics, Queen Mary, University of 
London, London E1 4NS, UK}
\vspace{0.2cm}

\vspace{0.2cm}
{\large G. Gerbier, J. Mallet, L. Mosca}

\vspace{0.2cm}
{\it DSM/DAPNIA/SPP, C.E.A. Saclay, F-91191 
Gif-sur-Yvette, France}
\vspace{0.2cm}

\vspace{0.2cm}
{\large C. Tao}

\vspace{0.2cm}
{\it CPPM, IN2P3/CNRS and Universit\'e Aix-Marseille II,
F-13288 Marseille, Cedex 09, France}
\vspace{0.2cm}

\begin{abstract}
The status of dark matter searches with inorganic
scintillator detectors at Boulby mine is reviewed and 
the results of tests with a CsI(Tl) crystal are
presented. The objectives of the latter experiment were to study
anomalous fast events previously observed 
and to identify ways to remove this background.
Clear indications were found that these events were due to
surface contamination of crystals by alphas, probably from
radon decay. A new array of unencapsulated NaI(Tl) crystals
immersed either in liquid paraffin or pure nitrogen atmosphere 
is under construction at Boulby. Such an approach allows
complete control of the surface of the crystals and 
the ability to remove any surface
contamination. First data from the
unencapsulated NaI(Tl) do not show the presence
of anomalous fast events.
\end{abstract}

\end{center}

\vspace{0.5cm}
\noindent {\it Key words:} Scintillation detectors, Inorganic crystals,
Dark matter, WIMP, Pulse shape analysis

\noindent {\it PACS:}  29.40.Mc, 14.80.Ly, 23.60.+e, 95.35.+d, 95.30.Cq

\vspace{0.5cm}
\noindent Corresponding author: V. A. Kudryavtsev, Department of Physics 
and Astronomy, University of Sheffield, Hicks Building, Hounsfield Rd., 
Sheffield S3 7RH, UK

\noindent Tel: +44 (0)114 2224531; \hspace{2cm} Fax: +44 (0)114 2728079; 

\noindent E-mail: v.kudryavtsev@sheffield.ac.uk
\pagebreak

{\large \bf 1. Introduction}
\vspace{0.3cm}

\indent The UK Dark Matter Collaboration (UKDMC) has been operating 
encapsulated NaI(Tl) detectors 
at the Boulby Mine 
underground site for several years \cite{UKDMC2}. 
Competitive limits on the flux of weakly interacting massive 
particles (WIMPs), that may 
constitute up to $90\%$ of mass in the Galaxy, have been set by 
one of these detectors using 
pulse shape analysis (PSA) to distinguish between
scintillation arising 
from background electron recoils and  
that due to nuclear recoils \cite{UKDMC1,UKDMC3}. 
Discrimination is possible because 
the sodium and iodine recoils expected 
from elastic scattering by WIMPs have faster mean pulse decay time 
than for electrons \cite{Dan1}. Traditionally, because NaI is 
hygroscopic, detectors are 
fabricated using an outer copper encapsulation with glued-in quartz 
windows plus additional thick 
(typically $>100$ mm) quartz lightguides to shield the crystal 
from photomultiplier activity.   
However, this design limits detector sensitivity because 
it prevents access to potential 
background sources on the crystal surfaces.
The importance of NaI(Tl) surfaces has been highlighted recently by 
indications that they might be a 
source of anomalous fast time constant events seen so far at similar
rates in many dark 
matter experiments with encapsulated
NaI(Tl) crystals \cite{UKDMC2,Saclay,VAK,Smith}. Greater 
access would allow improved control of potential contaminants there 
and hence a possible reduction 
in such events, leading to greater sensitivity to WIMPs.

In this paper we report the results of a study of anomalous fast 
events with NaI(Tl) and CsI(Tl) detectors and a recipe to suppress
their rate. We confirm also that unencapsulated
NaI(Tl) crystals normally do not have this background.

\vspace{0.5cm}
{\large \bf 2. Anomalous fast events}
\vspace{0.3cm}

\begin{figure}[htb]
\begin{center}
\epsfig{figure=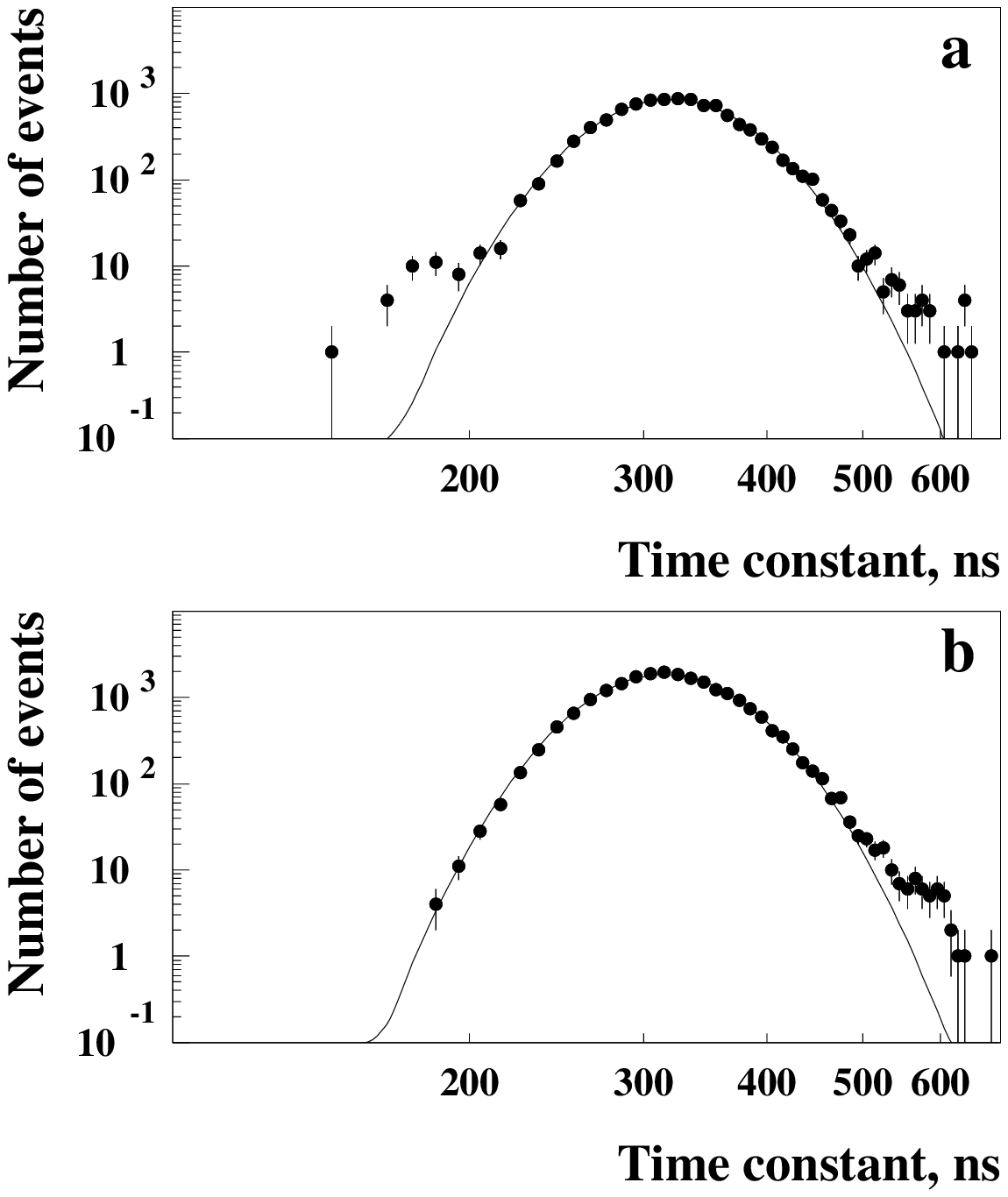,height=15cm}
\caption {a) Time constant distribution for events with visible energy
35-40 keV from one encapsulated NaI(Tl) detector; 
b) similar distribution for Compton events from a gamma source.
Solid curves show fits to Gaussian distributions
on a logarithmic scale ($\log$(Gauss)-function).}
\label{fig:bump46}
\end{center}
\end{figure}

\indent The observation of anomalous fast events in the UKDMC
encapsulated NaI(Tl) detectors was first reported by 
Smith {\it et al.} \cite{UKDMC2}. 
These events are faster than typical electron recoil pulses and
even faster than nuclear recoil pulses \cite{UKDMC2}.
Figure \ref{fig:bump46}a
shows typical time constant distribution of events from one
run with 330 kg $\times$ days exposure of a 5.2 kg NaI(Tl) 
encapsulated crystal. The time constant distribution
of events collected during calibration runs, when the crystal was
irradiated by photons from a $^{60}$Co source, is shown in Figure
\ref{fig:bump46}b for comparison. Both distributions are fitted to
a $\log$(Gauss) function, shown as a solid line in Figure \ref{fig:bump46}.
The full data analysis procedure is described elsewhere
\cite{Dan1,VAK,csi,naiad,Dan2}. A ``bump'' of anomalous fast events
is clearly seen on the left edge of the distribution shown in Figure
\ref{fig:bump46}a. Note the absence of the bump 
in Figure \ref{fig:bump46}b,
which implies that the bump is not due to any non-uniformity of
the crystal.

There is an excess of observed events over the $\log$(Gauss) fit
also at high values of time constant (see Figure \ref{fig:bump46}). 
This can be explained assuming stochastic pile-up of single thermal
photoelectrons \cite{Dan2}, occasional afterpulses and fluorescence
of the crystal after the scintillation pulse. 
The effect is seen for both ``data'' and ``calibration'' runs and 
does not interfere with the search for events faster than
electron recoil events, such as nuclear recoils from WIMP 
interactions with matter or other kinds of fast events. 

\begin{figure}[htb]
\begin{center}
\epsfig{figure=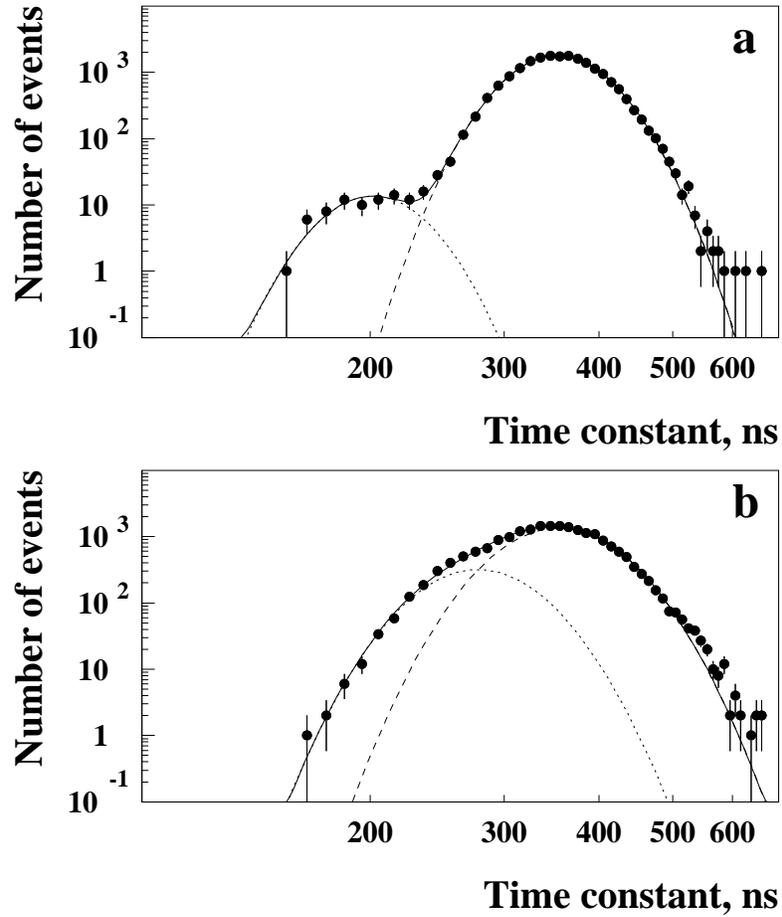,height=15cm}
\caption {a) Time constant distribution for events with visible energy
40-45 keV from encapsulated NaI(Tl) detector; 
b) similar distribution for calibration run with neutron source.
Solid curves show fits to a sum of 2 $\log$(Gauss)-functions.
Dashed curves show fits to gamma-induced events. Fits to
anomalous fast events (a) and neutron-induced events (b)
are plotted by dotted curves.}
\label{fig:ncal}
\end{center}
\end{figure}

Figure \ref{fig:ncal} show time constant distributions for another run with
the same encapsulated crystal (a) together with the results of a 
calibration run
with a neutron source (b). Both neutron-induced and gamma-induced events
are seen on the distribution plotted on Figure \ref{fig:ncal}b. 
The fits to gamma-induced events (electron recoils) are shown by
dashed curves. The fits to anomalous fast events (a) and neutron-induced
events (nuclear recoils) (b) are shown by dotted curves. 
From the comparison of Figures \ref{fig:ncal}a and \ref{fig:ncal}b 
we can conclude
that anomalous fast events are faster than nuclear recoil events,
expected from WIMP-nucleus interactions, and cannot be explained
by WIMPs or neutron background.

Similar fast events with comparable rate have been seen
also by the Saclay group \cite{Saclay}. One of the Saclay crystals
\cite{Saclay1}, of size, growth technique, manufacturer,
and housing similar to those used by the DAMA collaboration \cite{dama,dama1},
has been moved to Boulby (as a result of a collaboration between
UKDMC and Saclay) and is currently collecting data. We confirm
the presence of the population of fast events in this crystal
with a rate similar
to that seen in the UKDMC detectors (see also \cite{Saclay1,Spooner}
for discussion and Figure \ref{fig:bumplimits}).

Smith {\it et al}. \cite{UKDMC2} suggested that the anomalous fast events
could be due to MeV alphas. To account for the rate at low energies
(10-100 keV) the alphas would need to deposit a small fraction of their
initial energy at the crystal surface. Intrinsic bulk contamination
of the crystal by uranium and thorium (measured to be at the level of
about 0.1 ppb) is certainly not enough to explain the observed
high rate at
low energies. External incoming alphas from surrounding materials
(PTFE -- polytetrafluoroethylene, quartz windows) 
cannot easily explain the observed spectrum:
fine tuning of model parameters, such as a dead layer of scintillator,
is needed and a very high contamination of the material by uranium or
thorium (about 1 ppm) is required as well. 
Moreover, the time constant of the incoming alphas is not
matched well to that of the fast events \cite{VAK}.

Intrinsic surface contamination of the crystal by an alpha-emitting
isotope has recently been discussed as a source of the anomalous
fast events \cite{Smith}. Recoiling nuclei from radon decay
can be implanted into the crystal surface. This creates a thin
(0.1-0.2 microns) alpha emitting layer. Although a high
concentration of radioactive nuclei (0.1-1 ppm) is needed
to account for the observed rate, the predicted spectrum agrees
quite well with observations. Note that it is not known how such a
large concentration of radioactive nuclei can appear on the surface
of an encapsulated crystal.

Similar hypotheses on the source of the anomalous fast events
have been suggested by Saclay groups \cite{Saclay1,Saclay2} and
by Cooper {\it et al}. \cite{Cooper}. 
However, note that the hypothesis that anomalous events are due to
$^{214}$Po decay \cite{Cooper} requires a constant supply of radon
because of the short lives of isotopes decaying into $^{214}$Po.

If the source of fast events is indeed on the surface of the crystal, then
it can be removed by polishing the surface. This is hard to do with
encapsulated NaI(Tl) crystals but such an experiment can be done with 
CsI(Tl) providing it shows a similar rate of fast events. The
advantages of CsI(Tl) crystals are: a) they are only slightly 
hygroscopic and can be easily handled; b) they show better
discrimination capability between electron and nuclear recoils
\cite{csi,Pecourt}. (Note that CsI is generally ruled out for dark matter
searches due to high intrinsic background).

\vspace{0.5cm}
{\large \bf 3. Test with CsI(Tl) crystal}
\vspace{0.3cm}

\indent Tests were performed with an 0.8 kg CsI(Tl) Harshaw crystal 
previously studied in the 
laboratory to evaluate its characteristics such as
quenching factor for recoils and discrimination power, relevant to
dark matter searches. The results have been reported elsewhere
\cite{csi}. 

The crystal was subsequently moved to the underground laboratory at 
Boulby and tested for background rate and anomalous fast events.
In all tests we applied the standard procedure of pulse shape 
analysis adopted by the 
UK Dark Matter Collaboration for NaI(Tl) dark matter detectors 
\cite{UKDMC1,VAK,naiad,csi}. 
Pulses from both PMTs 
were integrated using a buffer circuit and then digitised using a 
LeCroy 9350A oscilloscope driven by 
a Macintosh computer running Labview-based data acquisition software. 
The digitised pulse shapes (10 $\mu$s digitisation time) 
were passed to the computer and stored on disk. Final analysis was 
performed on the sum of the pulses from the two PMTs. 
Our standard procedure of data analysis
involves the fitting of a single 
exponential to each integrated pulse to obtain the index of the exponent, 
$\tau$. Although scintillation pulses 
from CsI(Tl) have an additional second component \cite{csi}, 
the pulses can nevertheless be well 
fitted by a single exponential if fits are restricted to data below 
1500 ns. This fraction of 
the pulse contains the major contribution to the integrated pulse 
amplitude so that the distortion of the 
fit due to the presence of the second exponential at large time scales 
was found to be insignificant \cite{csi}.
This approximation has the advantage that a three parameter fit can be 
used on each pulse and a simple 
discrimination parameter defined, rather than a considerably more 
complicated six parameter fit in the 
case of two decay constants. The free fit parameters used are: 
the time constant of the single exponent, 
$\tau$; a normalisation constant and the start time of the pulse.

For each run the distribution of the 
number of events 
versus the time constant of the exponent ($\tau$) was generated for a 
range of energy bins. $\tau$-distributions 
for each population of pulses can be approximated by a gaussian in 
$\ln(\tau)$ ($\log$(Gauss) function) \cite{UKDMC1,VAK} (for a more detailed 
discussion of the distributions see \cite{Dan2} and references therein):

\begin{equation}
{{dN}\over{d\tau}} = {{N_o}\over{\tau \sqrt{2\pi} \ln w}} \times
\exp \Big[ {{-(\ln \tau-\ln \tau_o)^2}
\over{2(\ln w)^2}} \Big]
\end{equation}

The CsI(Tl) $\tau$-distributions were fitted with this 
gaussian in $\ln(\tau)$ 
with the three free parameters: time constant
$\tau_o$, width $w$ and normalisation factor $N_o$. 
In experiments where a second population
is seen (for example, nuclear recoils from a neutron 
source or anomalous fast events), the resulting 
$\tau$-distribution can be fitted with 
two $\log$(Gauss) functions with the same width $w$. 

\begin{figure}[htb]
\begin{center}
\epsfig{figure=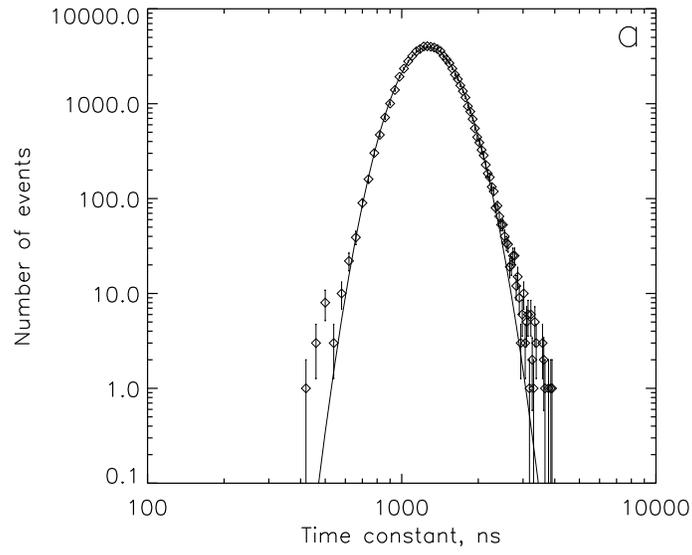,height=8cm}
\epsfig{figure=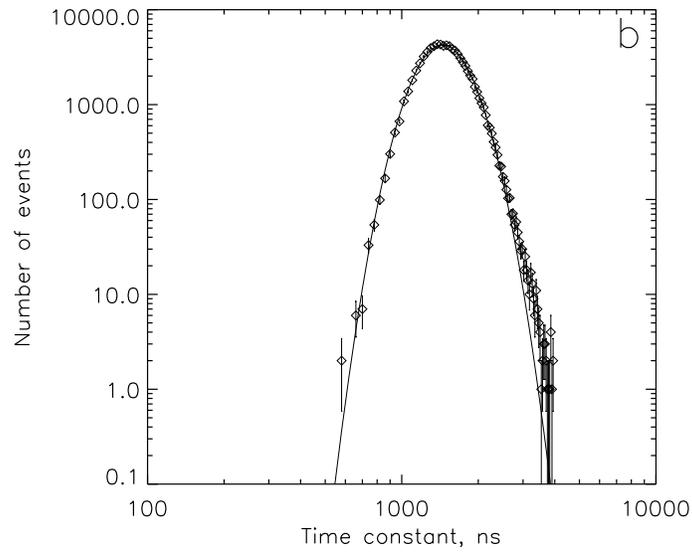,height=8cm}
\caption {a) Time constant distribution for events with visible energy
30-50 keV from the CsI(Tl) crystal; 
b) similar distribution after crystal polishing.
Solid curves show fits to a $\log$(Gauss)-function.}
\label{fig:csibump}
\end{center}
\end{figure}

\begin{figure}[htb]
\begin{center}
\epsfig{figure=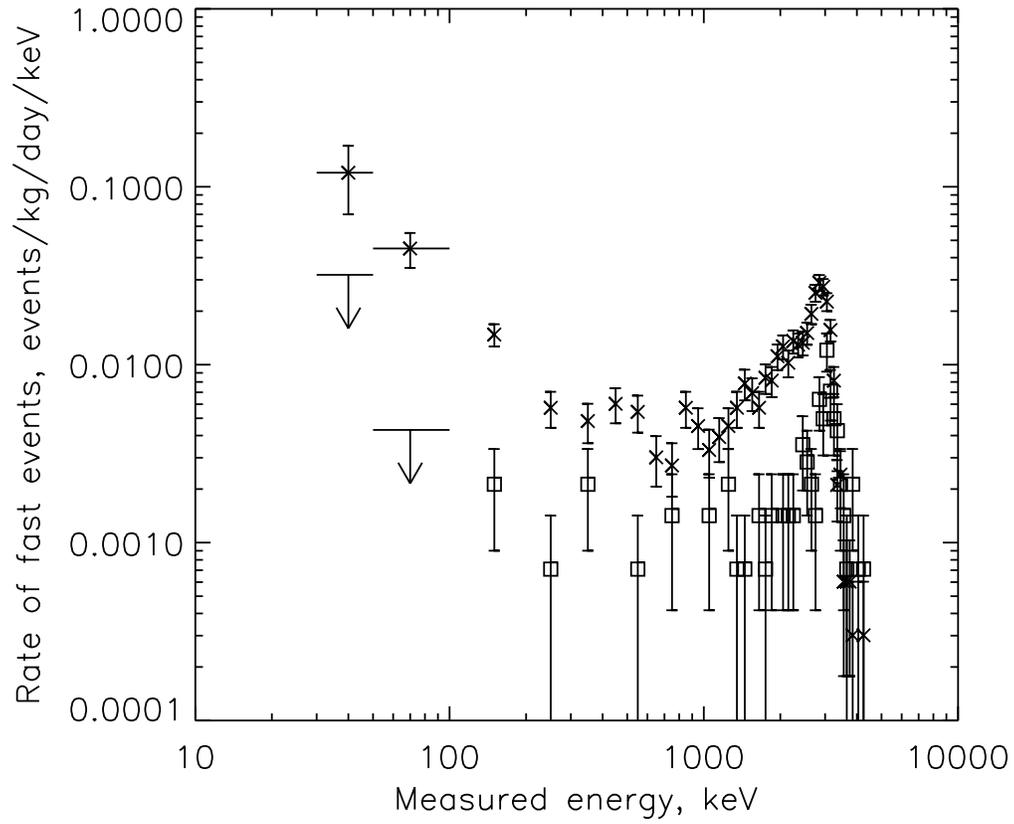,height=12cm}
\caption {Rate of fast events ($\alpha$s) in CsI(Tl) crystal
before (crosses) and after (squares) polishing. The first
two points after polishing show limits at 90\% confidence level.}
\label{fig:csisp}
\end{center}
\end{figure}

In the low background conditions of the
underground laboratory at Boulby, the CsI(Tl) crystal was found to show
an anomalous population of fast events.
Figure \ref{fig:csibump}a shows time constant distribution
of events with visible energy 30-50 keV
together with a fit to a $\log$(Gauss) function (14.3 kg $\times$ days 
of exposure). The spectrum
of these events  presented
in Figure \ref{fig:csisp} (crosses) has been traced up to MeV energies.
The rate and shape of the spectrum
are similar to those observed in the
NaI(Tl) encapsulated detectors \cite{UKDMC2}. 
A peak at about 2.9 MeV corresponds to the 5.3 MeV alphas from 
$^{210}$Po $\rightarrow$ $^{206}$Pb $\alpha$-decay, assuming
a quenching factor for alphas of 0.55. (Scintillation
efficiency of $0.50 \pm 0.05$ for alphas was measured for this crystal
with $^{241}$Am and $^{137}$Cs sources). Note the absence of
higher energy events which may be associated with the decay
channels prior to $^{210}$Po. This is in contradiction to what was 
suggested in Ref. \cite{Cooper}.

After 2 months of running at various dynamic ranges the crystal
was removed, polished and put into
a sealed vessel with nitrogen atmosphere. After polishing the crystal was
exposed to air for only a few hours during the installation procedure.

The subsequent runs revealed a decrease in the rate of fast
events by about a factor of 4 (squares in Figure \ref{fig:csisp}). 
The first two
points below 100 keV show an upper limit to the rate. An accurate 
measurement of the rate at these energies is difficult because of
the small mass of the crystal and the high rate of $\gamma$-background 
observed due to internal contamination of
$^{137}$Cs. The time constant distribution for events of 30-50 keV after
polishing is shown in Figure \ref{fig:csibump}b (22.3 kg $\times$ days 
of exposure). It can be seen that
the rate of anomalous fast events is significantly reduced
(see Figure \ref{fig:csibump}a), though not completely suppressed
probably due to the difficulty of removing the hard surface 
layer of CsI.

17 high-energy events (visible energy of 5-6 MeV) were also
detected during the first day after polishing. These are
double-pulse events where the first pulse corresponds to the
$\beta$-decay of $^{212}$Bi (half-life 1 hour) 
and the second due to the
$\alpha$-decay of $^{212}$Po (half-life 0.3 $\mu$s). These
events are probably caused by contamination of the crystal
surface during installation. No more of these events were
seen after the first day (half-life of parent isotope
$^{212}$Pb is 10.6 hours). Visible energy of these double-pulse
events agrees with the assumption that the scintillation 
efficiency of alphas is about 0.5.

Only one prominent peak is seen in the spectra shown in
Figure \ref{fig:csisp}. The peak is probably due to $^{210}$Po 
$\rightarrow$ $^{206}$Pb $\alpha$-decay (5.3 MeV $\alpha$s).
No decay chains 
$^{222}$Rn $\rightarrow$ $^{218}$Po $\rightarrow$ $^{214}$Pb
or
$^{224}$Ra $\rightarrow$ $^{220}$Rn $\rightarrow$ $^{216}$Po
$\rightarrow$ $^{212}$Pb were seen before or after
polishing. This suggests that the concentration of U, Th and
Ra in the bulk of the crystal is very low (less than 0.1 ppb).

At least several months of exposure to Rn is needed to explain the 
rate of $\alpha$-events in CsI(Tl) and NaI(Tl) detectors.
This is not surprising for an unencapsulated CsI crystal but
is hard to explain for NaI sealed detectors.

\vspace{0.5cm}
{\large \bf 4. Array of unencapsulated NaI(Tl) detectors - NAIAD}
\vspace{0.3cm}

\indent The results obtained with the CsI(Tl) crystal at Boulby clearly
indicate the importance of having access to the crystal surface
for polishing and control. Such access can be granted by
running unencapsulated crystals in high purity mineral oil
or dry gas inside
sealed plastic or copper vessels. Laboratory tests have shown
also that high light yield (up to 10 photoelectrons per keV)
can be reached with the aforementioned detectors \cite{naiad,Peak}.

Since 1999 the UKDMC has been developing a programme to run several
unencapsulated NaI(Tl) crystals mounted in an array -- NAIAD
(NAI Advanced Detector). The NAIAD array is designed to be
flexible enough to allow various modes of operation with
crystals. To date two types of
module have been constructed: a ``vertical'' module filled
with high purity mineral oil to protect crystals from moisture,
and a ``horizontal'' module in which either an oil or
dry nitrogen is used around the crystal. Operation of an
encapsulated crystal is also possible in the horizontal module.
Design details and predictions of sensitivity to WIMPs are
given in Ref. \cite{naiad}.

\begin{figure}[htb]
\begin{center}
\epsfig{figure=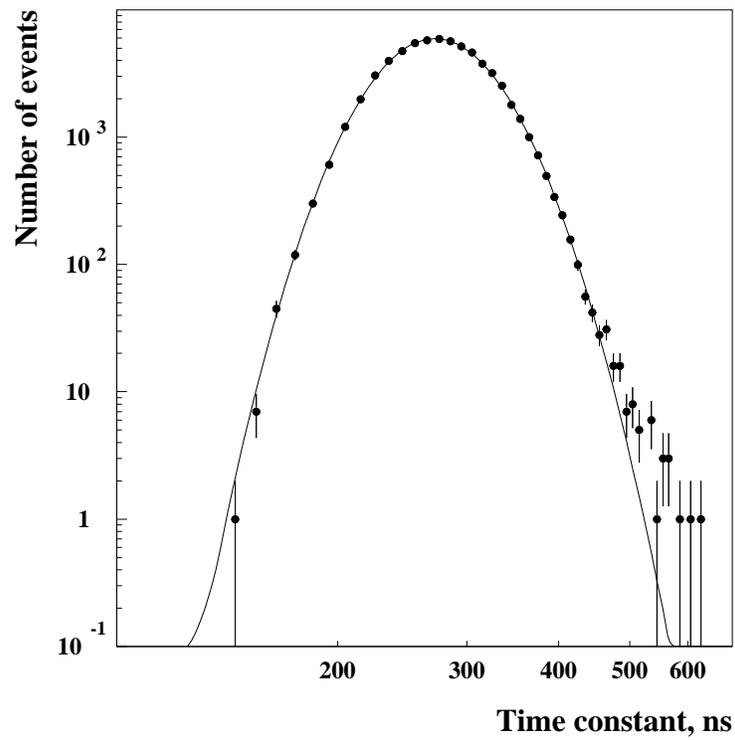,height=12cm}
\caption {Time constant distribution for events with visible energy
35-40 keV from the first NAIAD module (DM74).
Solid curve shows a fit to a $\log$(Gauss)-function.}
\label{fig:tau}
\end{center}
\end{figure}

\begin{figure}[htb]
\begin{center}
\epsfig{figure=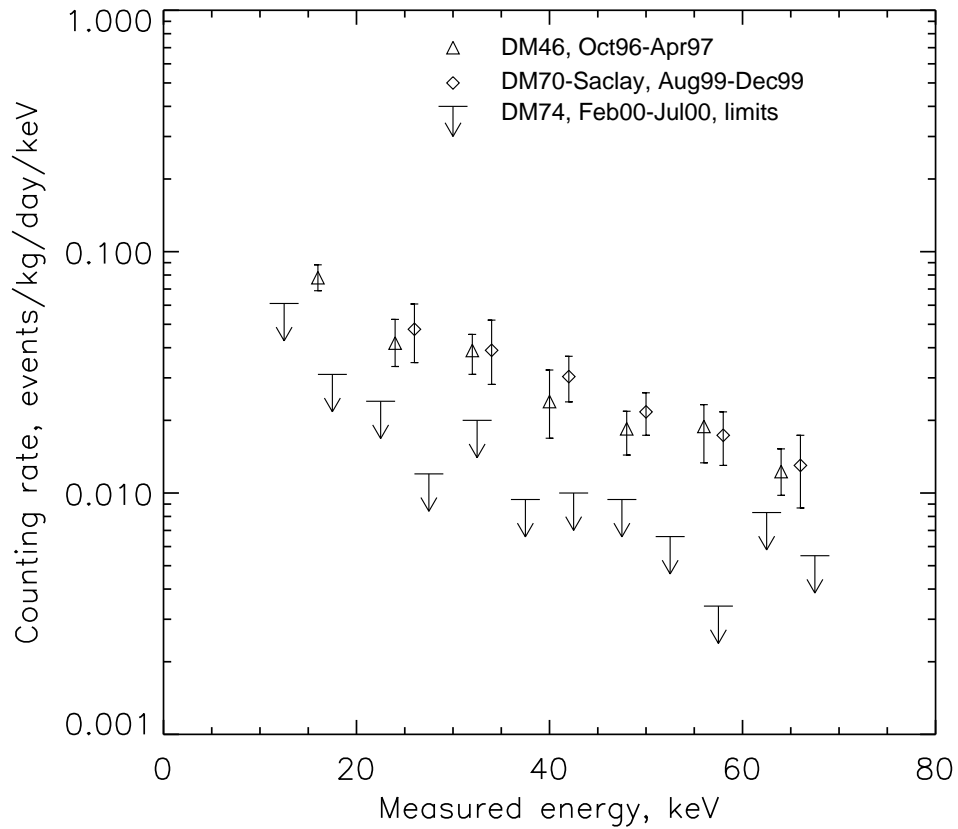,height=12cm}
\caption {Rate of anomalous fast events in NaI(Tl) 
encapsulated crystals and limits on the rate of these
events in the unencapsulated crystal (DM74):
triangles -- UKDMC DM46 crystal;
diamonds -- Saclay crystal, also named DM70 (see also
\cite{Saclay1,Spooner}); arrows -- 
limits at 90\% confidence level from DM74 data.}
\label{fig:bumplimits}
\end{center}
\end{figure}

The first vertical module of NAIAD has been running since February
2000. It contains a 14 cm diameter $\times$ 15 cm length crystal
(termed DM74) with mass of about 8.5 kg. The crystal was polished before 
installation. The total exposure (excluding
calibration runs) is 996 kg $\times$ days.
Figure \ref{fig:tau} shows a typical distribution of time constants
for events with 35-40 keV visible energy together with a fit to
a log(Gauss) function. 
As with the polished CsI(Tl) crystal the rate of anomalous fast events
is greatly suppressed.
The calculated limit on the rate of anomalous
fast events as a function of visible energy is shown in 
Figure \ref{fig:bumplimits} together with the energy spectra of fast
events measured in typical encapsulated crystals.

A second (horizontal) module containing
4 kg unencapsulated and polished crystal in nitrogen (DM72) also 
does not show
the presence of fast events. This second detector is currently
running underground at Boulby. Analysis of data
from both crystals in terms of limits on the WIMP-nucleon
and WIMP-proton cross-sections is in progress.

\vspace{0.5cm}
{\large \bf 5. Pulse shape analysis versus annual modulation}
\vspace{0.3cm}

\indent Pulse shape analysis (PSA) is not the only technique used 
with NaI(Tl) detectors for dark matter searches. The DAMA group \cite{dama}
searches for an annual modulation in the background counting rate
of their NaI(Tl) array without PSA 
(PSA is used only to discriminate between scintillation
pulses and PMT noise). Evidence for such an annual modulation in the 
background
rate has been reported by DAMA (see \cite{dama} and references
therein) indicating a possible signal from WIMPs. 

It would appear that the DAMA group could, in fact, 
confirm or exclude this possibility 
using pulse shape analysis. In Ref. \cite{dama} DAMA
presented results of the annual modulation analysis 
(positive signal) together with previous limits on WIMP-nucleon 
cross-section obtained with PSA. All sets of data (DAMA/NaI-0 analysed
using PSA and DAMA/NaI-1 -- DAMA/NaI-4 analysed using annual modulation) 
were obtained with the same experimental set-up and under 
similar conditions such as background rate etc. Pulse shape analysis 
applied to the first
data set (DAMA/NaI-0) allowed DAMA to put limits on WIMP-nucleon 
interactions of the order of $(5-6) \times 10^{-6}$ pb for 50-100 GeV
mass WIMPs with halo density of 0.3 GeV/cm$^{3}$ 
(see \cite{dama,dama1} for details and halo parameters used).
This limit was obtained with 4123.2 kg $\times$ days exposure.
The subsequent data sets (DAMA/NaI-1 -- DAMA/NaI-4) 
totaling 57986 kg $\times$ days showed a positive signal
at 4 $\sigma$ confidence level
using annual modulation analysis without PSA. WIMP parameters derived
from this analysis are:
$M_W=(52^{+10}_{-8})$ GeV and cross-section 
$\sigma_p=(7.2^{+0.4}_{-0.9}) \times 10^{-6}$ pb 
with the same halo parameters \cite{dama}. Simple
statistical considerations show that 
PSA applied to this 15 times larger exposure
from all five data sets (DAMA/NaI-0 -- DAMA/NaI-4) compared to the first 
one (DAMA/NaI-0) could yield a limit on the WIMP-nucleon cross-section
improved by a factor of 3.9.
Such an analysis would allow DAMA either to confirm
the modulated signal or to exclude practically 
the whole region of parameters that they derive from the observed
modulated signal (see figures 3 and 4 in Ref.
\cite{dama} for the DAMA allowed region of WIMP parameters).
The same statistical considerations suggest that the signal reported in 
Ref. \cite{dama2} for the period DAMA/NaI-1 with 4549 kg $\times$ days
should have been at a 1.1 $\sigma$ confidence level, to be consistent
with the signal observed at 4 $\sigma$ confidence level with
57986 kg $\times$ days exposure.

\vspace{0.5cm}
{\large \bf 6. Conclusions}
\vspace{0.3cm}

\indent Tests with a CsI(Tl) and unencapsulated NaI(Tl) crystals
have shown that anomalous fast events
seen in several NaI(Tl) detectors at Boulby were probably
due to surface $\alpha$s. Radioactive $\alpha$-emitting isotopes
had likely been implanted into crystal surfaces by radon decay.
Polishing the crystal surfaces removed a major part of the fast
events. A new array of unencapsulated NaI(Tl) crystals (NAIAD)
is being installed in the underground laboratory at Boulby.
Data from the first of these modules do not reveal anomalous fast events.

\vspace{0.5cm}
{\large \bf Acknowledgments}
\vspace{0.3cm}

\indent The Collaboration wishes to thank PPARC for financial support.
We are also grateful to the staff of Cleveland Potash Ltd. for assistance.

\pagebreak

\pagebreak

\pagebreak

\pagebreak

\pagebreak

\pagebreak

\end{document}